\pdfoutput=1
\RequirePackage{fix-cm}
\documentclass[11pt, tightenlines, floatfix,pra,twocolumn,secnumarabic,amssymb,amsmath,nobibnotes,aps,prl,superscriptaddress,preprintnumbers]{revtex4-1}

\usepackage[fontsize=10.25pt]{scrextend}

\usepackage{booktabs}
\usepackage[margin=2cm]{geometry}
\usepackage{caption}
\usepackage{url}
\usepackage{subcaption}
\PassOptionsToPackage{hyphens}{url}\usepackage{hyperref}
\usepackage[hyphens]{url}
\usepackage{graphicx}
\usepackage{dcolumn}
\usepackage{bm}
\usepackage{makecell} 
\usepackage{color}
\usepackage{breakcites}
\usepackage{ragged2e}
\usepackage{enumitem}

\newlength\lengtha 
\newlength\lengthb 

\DeclareCaptionJustification{justified}{\justifying}
\usepackage[normalem]{ulem} 

\usepackage[utf8]{inputenc}
\usepackage[T1]{fontenc}
\usepackage{mathptmx}

\begin{document}

\title[]{Chemical diversity in molecular orbital energy predictions with kernel ridge regression}

\author{Annika Stuke}
 \affiliation{Department of Applied Physics, Aalto University, P.O. Box 11100, Aalto FI-00076, Finland}
  \email{annika.stuke@aalto.fi}
  
  \author{Milica Todorovi\'{c}}%

\affiliation{Department of Applied Physics, Aalto University, P.O. Box 11100, Aalto FI-00076, Finland
}%

 \author{Matthias Rupp}
\affiliation{%
Fritz Haber Institute of the Max Planck Society, Faradayweg 4-6, 14195 Berlin, Germany}

\author{Christian~Kunkel}
\affiliation{Chair for Theoretical Chemistry and Catalysis Research Center, Technische Universit{\"a}t M{\"u}nchen, Lichtenbergstr. 4, 85747 Garching, Germany}
\affiliation{Department of Applied Physics, Aalto University, P.O. Box 11100, Aalto FI-00076, Finland}

\author{Kunal Ghosh}%

\affiliation{Department of Applied Physics, Aalto University, P.O. Box 11100, Aalto FI-00076, Finland
}%
\affiliation{Department of Computer Science, Aalto University, P.O. Box 15400, Aaalto FI-00076, Finland
}%

\author{Lauri Himanen}
 \affiliation{Department of Applied Physics, Aalto University, P.O. Box 11100, Aalto FI-00076, Finland}
\author{Patrick Rinke}
 \affiliation{Department of Applied Physics, Aalto University, P.O. Box 11100, Aalto FI-00076, Finland}
 \affiliation{Chair for Theoretical Chemistry
 and Catalysis Research Center, Technische Universit{\"a}t M{\"u}nchen, Lichtenbergstr. 4, 85747 Garching, Germany}

\date{\today}

\begin{abstract}
Instant machine learning predictions of molecular properties are desirable for materials design, but the predictive power of the methodology is mainly tested on well-known benchmark datasets. Here, we investigate the performance of machine learning with kernel ridge regression (KRR) for the prediction of molecular orbital energies on three large datasets: the standard QM9 small organic molecules set, amino acid and dipeptide conformers, and organic crystal-forming molecules extracted from the Cambridge Structural Database. We focus on prediction of highest occupied molecular orbital (HOMO) energies, computed at density-functional level of theory. Two different representations that encode molecular structure are compared: the Coulomb matrix (CM) and the many-body tensor representation (MBTR). We find that KRR performance depends significantly on the chemistry of the underlying dataset and that the MBTR is superior to the CM, predicting HOMO energies with a mean absolute error as low as 0.09\,eV. To demonstrate the power of our machine learning method, we apply our model to structures of 10k previously unseen molecules. We gain instant energy predictions that allow us to identify interesting molecules for future applications.
\end{abstract}

\maketitle

\section{\label{sec:introduction}Introduction}

Machine learning (ML) in molecular and materials science has gained increased attention in the last decade and its application domain is widening continuously.\cite{rlb2018} Applications include the search for improved and novel materials, \cite{Rampi_review,Zunger:2018} computational drug design, \citep{ma_deep_2015} battery development, \citep{sendek_machine_2018, shandiz_application_2016} identification of new molecules for organic light-emitting diodes \citep{gomez_design_2016} or catalyst advancements for greenhouse gas conversion. \citep{goldsmith_machine, meyer_machine_2018} In the context of \textit{ab initio} molecular science, ML has been applied to learn a variety of molecular properties such as atomization energies, \citep{hansen_assessment_2013, rupp_fast_2012, huang_communication_2016, faber_prediction_2017, faber_alchemical_2018, bartok_machine_2017, collins_constant_2018, de_comparing_2016, ramakrishnan_big_2015, montavon_machine_2013} polarizabilities, \citep{faber_alchemical_2018, schutt_schnet_2018, huang_communication_2016, collins_constant_2018, montavon_machine_2013} electron ionization energies and affinities, \citep{huang_communication_2016, collins_constant_2018, montavon_machine_2013,Gosh/etal:2019} dipole moments, \citep{pereira_machine_2018, faber_prediction_2017, faber_alchemical_2018, schutt_schnet_2018, huang_communication_2016,bal2015bq,bdtl2018q} enthalpies, \citep{faber_prediction_2017, huang_communication_2016, ramakrishnan_big_2015} band gaps, \citep{faber_prediction_2017, faber_alchemical_2018, schutt_schnet_2018, huang_communication_2016}, binding energies on surfaces \cite{Todorovic/etal:2019} as well as heat capacities. \citep{faber_prediction_2017, faber_alchemical_2018, huang_communication_2016} A few studies addressed the prediction of spectroscopically relevant observables, such as electronic excitations, \citep{huang_communication_2016, collins_constant_2018, montavon_machine_2013, pronobis_capturing_2018} ionization potentials, \citep{huang_communication_2016, collins_constant_2018, montavon_machine_2013, ramakrishnan_electronic_2015} nuclear chemical shifts, \citep{rupp_machine_2015-1} atomic core level excitations \citep{rupp_machine_2015-1} or forces on atoms. \citep{rupp_machine_2015-1}
In this paper, we employ kernel ridge regression (KRR) to predict the energy of the highest occupied molecular orbital (HOMO), which is of particular current interest for the development of new substances and materials. Electronic devices based on organic compounds are widely used in the technological industry and frontier orbital energies give important information about the optoelectronic properties of possible candidate materials. 

While the majority of ML studies focuses on ground state properties, a handful of studies have addressed the pre-
\begin{figure}[t] 
\centering
    \includegraphics[width=8.4cm]{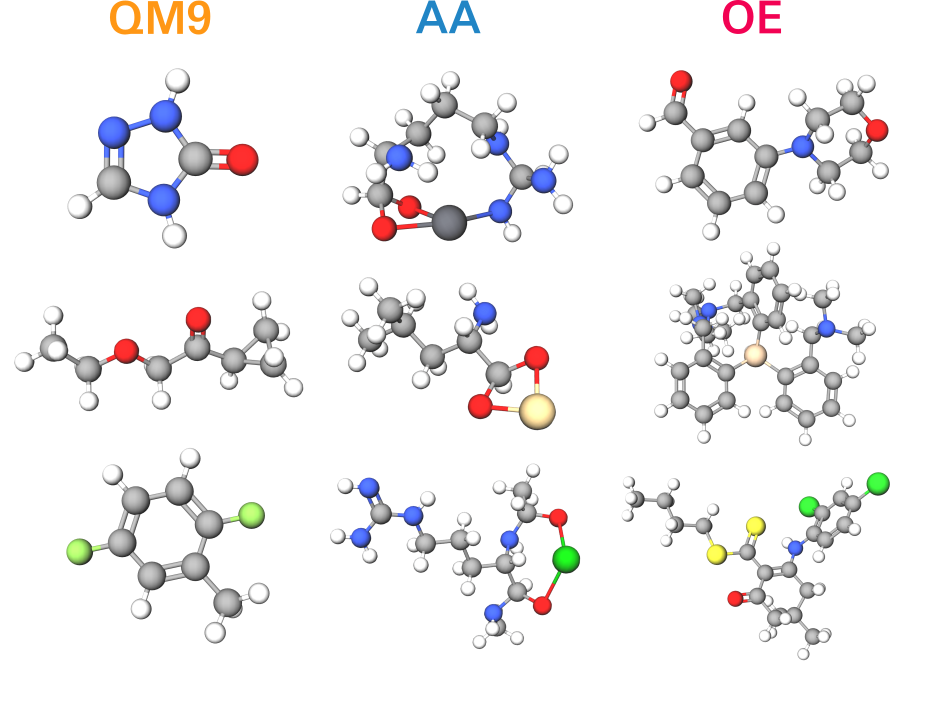}
    \caption{\label{fig:molecules} Example molecules taken from the three datasets used in this work (see Appendix \ref{sec:chem_names} for chemical names). 
    Depicted elements are H (white), C (grey), N (blue), F (light green), O (red), Pb (dark grey), Cd (gold), Ba (dark green), Si (bronze), Cl (green) and S (yellow).}
    \vspace{-0.5cm}
\end{figure}
diction frontier orbital energies, for example using neural networks, \citep{montavon_machine_2013, pyzer_learning_2015, schutt_schnet_2018, faber_prediction_2017,Gosh/etal:2019} random forest models, \citep{pereira_machine_2017} or kernel ridge regression. \citep{huang_communication_2016, collins_constant_2018, faber_alchemical_2018, de_comparing_2016} To our knowledge, the best prediction accuracy reported for HOMO energy predictions on the well-known QM9 dataset \citep{ramakrishnan_quantum_2014} of small organic molecules was achieved with a deep neural network and featured a mean absolute error (MAE) of 0.041 eV. \citep{schutt_schnet_2018} QM9 was also employed by many other studies to explore the effect of molecular descriptors on prediction accuracy. \citep{rupp_fast_2012,huang_communication_2016,faber_prediction_2017,schutt_schnet_2018}

While these results are quantitative and valuable, it is not clear to what extent the predictive power of the employed methodology is transferable to other molecular datasets. Motivated by optoelectronic applications, we are interested in HOMO predictions for large optically-active molecules with complex aromatic backbones and diverse functional groups which differ notably from the QM9 dataset.

We here employ KRR on three different datasets, two of which have not yet been used for the prediction of molecular orbital energies with ML. These two less-known datasets consist of 44\,k conformers of proteinogenic amino acids from a public database of oligo-peptide structures \citep{ropo_first_2016} and 64\,k organic molecules extracted from the organic crystals of the Cambridge Structural Database.\citep{Schober/etal:2016,groom_cambridge_2016} In addition, we also use the well-known QM9 benchmark database of 134\,k small organic molecules \citep{ruddigkeit_enumeration_2012, ramakrishnan_quantum_2014} as a third dataset to compare results with previous studies in this field on a common basis. For all three datasets, we have calculated reference HOMO energies with density functional theory (DFT). 

Moreover, we compare the performance of two different molecular representations: the well-studied Coulomb matrix (CM), which is simple, easy to compute and yields fast and inexpensive ML predictions. The second is a constant-size representation recently introduced by one of us \citep{huo_unified_2017} that relies on interatomic many-body functions including bonding and angular terms, called the many-body tensor representation (MBTR). While previous studies already demonstrated that the CM can easily be outperformed by more sophisticated molecular descriptors \cite{huang_communication_2016,collins_constant_2018,faber_prediction_2017}, we aim to analyze the degree of accuracy that can be achieved with the simple and cheap CM in comparison to the costlier MBTR.  

The primary goal of our study is the comparison of KRR performance across three datasets with different chemical diversity. We show that the accuracy of HOMO energy predictions with KRR depends -- besides the choice of molecular representation -- crucially on the chemistry of the underlying dataset. Our measurable acceptable accuracy for HOMO energy predictions is 0.1 eV. Experiments for HOMO energy determination typically have a resolution of several tenth of eV, and prediction errors of state-of-the-art theoretical spectroscopy methods commonly range between 0.1 and 0.3 eV. We demonstrate how differences in model performance across chemically diverse settings can be related to certain dataset properties. Moreover, we quantify molecular orbital energy predictions that are presently available for realistic datasets of technological relevance.

Once trained, our KRR model can make instant HOMO energy predictions for numerous unknown molecules at no further cost. We demonstrate this by producing a spread of HOMO energy predictions for a new dataset of 10k organic molecules \cite{ramakrishnan_big_2015}, whose original HOMO energies are unknown. With instant energy predictions for all 10k molecules we gain a rough estimate of the HOMO energy distribution for this dataset. We can further identify interesting molecules within a certain energy range for additional analysis. Hence, large numbers of new molecules, whose orbital energies have not yet been measured or computed, can be quickly screened for their usability in future applications. In this way, KRR can complement conventional theoretical and experimental methods to greatly accelerate the analysis of materials.

The manuscript is organized as follows:
In Section~\ref{sec:datasets} we introduce the three datasets used in this work. In Section~\ref{sec:representation} we briefly review the two descriptors that we employ in our ML framework, which is described in Section~\ref{sec:method}. In Section~\ref{sec:results} we present our results and then discuss our findings in Section~\ref{sec:discussion}.

\section{\label{sec:datasets}Datasets}

We train and evaluate our model on three different datasets, for which example molecules are depicted in Fig.~\ref{fig:molecules}. 
For all three datasets, we performed DFT calculations with the FHI-aims code. \citep{blum_ab_2009,HavuV09,Levchenko15,Xinguo/implem_full_author_list} We optimized the atomic structure of all molecules using the Perdew-Burke-Ernzerhof (PBE) functional \citep{Perdew1996} including Tkatchenko-Scheffler van der Waals corrections (PBE+vdW), \citep{Tkatchenko2009}  tight computational settings and the tier 2 basis sets of FHI-aims. For reasons of computational tractability, we also calculate HOMO energies with DFT, by taking the eigenvalue of the highest occupied molecular state from the PBE+vdW calculation. 
\begin{figure}[h!]
     \centerline{\includegraphics[width=8.45cm]{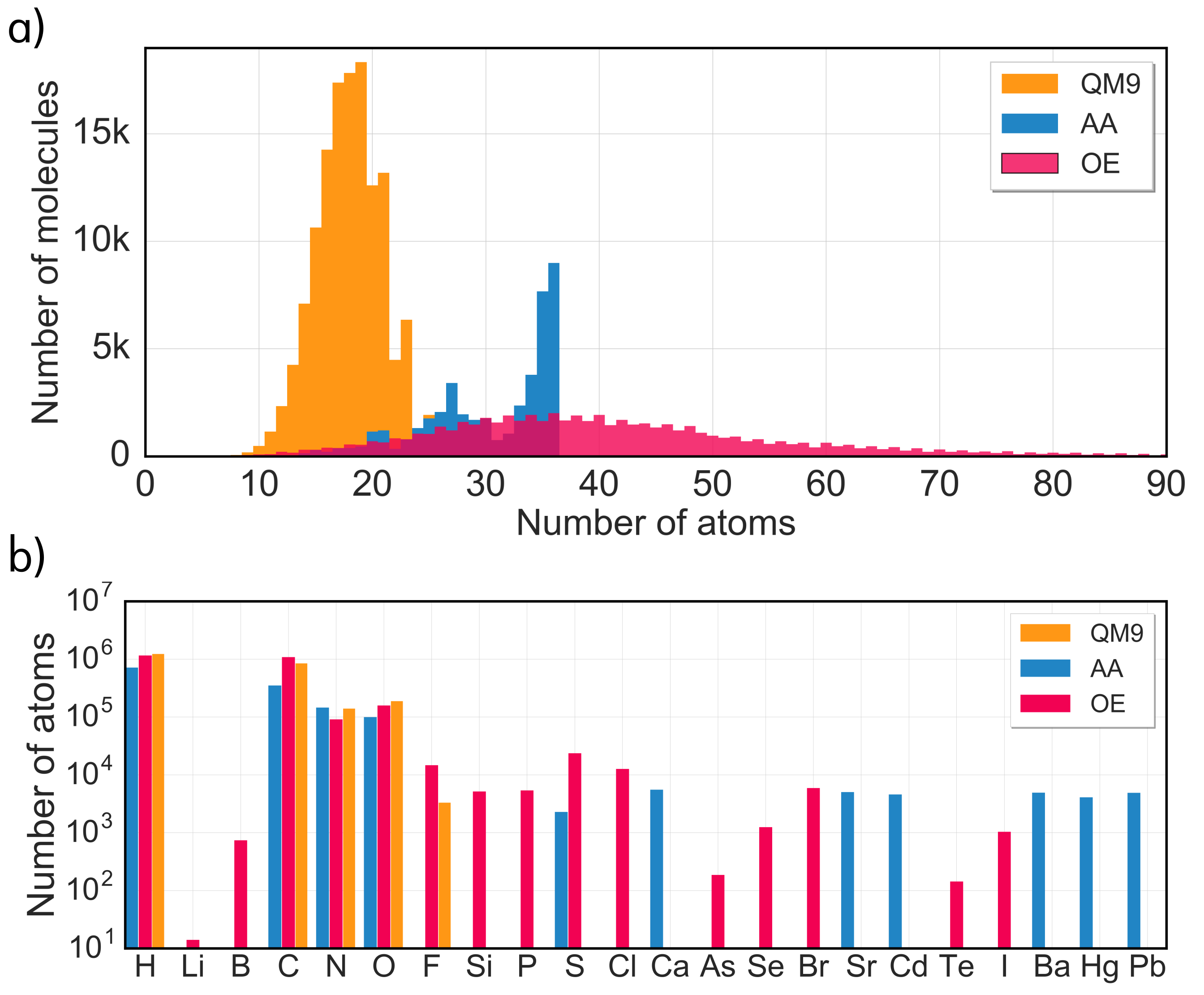}}
\caption{\label{fig:statistics}Distribution of a) molecular size (including H atoms) and b) element type within QM9 (orange), AA (blue) and OE (red). OE molecules reach a size of up to 174 atoms. Since the occurence of large molecules with more than 90 atoms is very low (they amount to 1.4\% of the entire dataset), we cut off the distribution at 90 atoms to facilitate a better comparison with the other two datasets. 
}
 \end{figure}
 
\begin{figure}[t!] 
\centerline{\includegraphics[width=9cm]{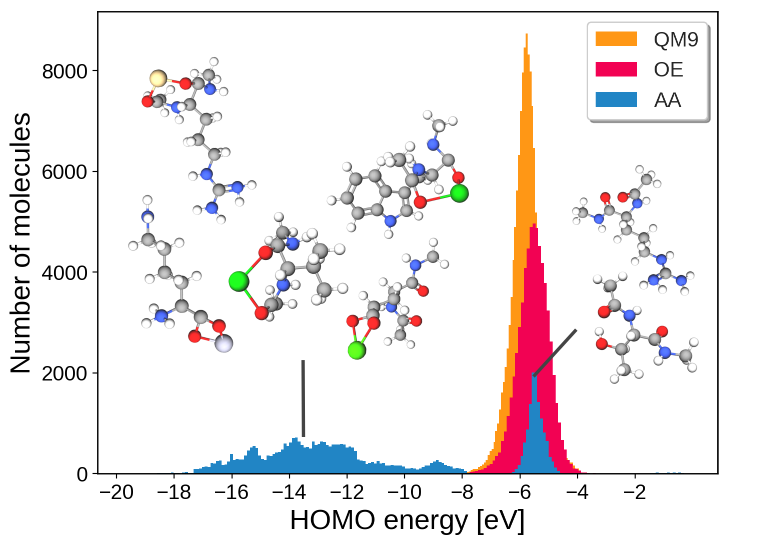}}
\caption{\label{fig:homo}Distributions of pre-computed HOMO energies for all three datasets. In QM9 and OE, HOMO energies are centered around -6 eV, while most HOMO energies in AA are distributed over a wider negative range. Shown as an inset are example molecules from the AA dataset and their location in the HOMO energy distribution. The HOMO energies in the deep negative range correspond to amino acids and dipeptides with one of six different metal cation additions: Ca$^{2+}$, Ba$^{2+}$, Sr$^{2+}$, Cd$^{2+}$, Pb$^{2+}$ or Hg$^{2+}$. The HOMO energies centered around -6 eV correspond to bare amino acids and dipeptides.}
\end{figure}
\begin{center}
\begin{table}[h!]
\setlength{\tabcolsep}{4pt}
\begin{tabular}{c|cccc}
\hline
\hline
\textbf{Dataset} & \textbf{Mean [eV]}  & \textbf{Std. dev. [eV]} & \textbf{Min [eV]}  & \textbf{Max [eV]}  \\
\hline
QM9 & -5.77& 0.52 &-10.45 & -2.52 \\
AA  & -10.85 & 3.97 & -19.63 & -0.06 \\
OE  & -5.49 & 0.57 & -12.70 & -2.73 \\
\hline
\hline
\end{tabular}
\captionof{table}{Mean value, standard deviation and ranges of DFT-computed HOMO energies for all three datasets as depicted in Fig. \ref{fig:homo}. QM9 and OE have similar mean values, ranges and standard deviations of HOMO energies, while the HOMO energies of the AA dataset spread out over a wide range of values due to metal cations with large atomic number that are attached to the amino acids and dipeptides.} \label{table:homo}
\end{table}
\end{center}
\vspace{-2cm}
Although DFT Kohn-Sham energies have limited accuracy, they provided us with a large, convenient and consistent dataset for developing our methodology. In the future, we will extend our study to HOMO energies computed with the more appropriate $GW$ method.\cite{Hedin:1965,Rinke/etal:2005} However, at present it is not possible to calculate hundreds of thousands of molecules with the $GW$ method with reasonable computational resources. In the following, we describe the datasets in more detail.

\subsection{\label{sec:134k}QM9: 134\,k small organic compounds}
This dataset is extracted from the QM9 database and consists of 133,814 small organic molecules with up to 9 heavy atoms made up of C, N, O and F atoms \citep{ramakrishnan_quantum_2014}. 
It contains small amino acids and nucleobases, pharmaceutically relevant organic building blocks, for a total of 621 stoichiometries. The QM9 database has been used in a variety of ML studies \citep{collins_constant_2018, huang_communication_2016, bartok_machine_2017, ramakrishnan_big_2015, ramakrishnan_machine_2017, schutt_schnet_2018, faber_prediction_2017, pronobis_capturing_2018, schutt_quantum-chemical_2017, faber_alchemical_2018, lubbers_hierarchical_2018, unke_a_2018} and has become the \emph{drosophila} of ML in chemistry.

\subsection{\label{sec:AA}AA: 44\,k amino acids and dipeptides}
This dataset, denoted AA, contains 44,004 isolated and cation-coordinated conformers of 20 proteinogenic amino acids and their amino-methylated and acetylated (capped) dipeptides. \citep{ropo_first_2016} The molecular structures are made of up to 39 atoms including H, C, N, O, S, Ca, Sr, Cd, Ba, Hg and Pb. The amino acid conformers reveal different protonation states of the backbone and the sidechains. Furthermore, amino acids and dipeptides with divalent cations (Ca$^{2+}$, Ba$^{2+}$, Sr$^{2+}$, Cd$^{2+}$, Pb$^{2+}$, and Hg$^{2+}$) are included. Since all amino acids share a common backbone the complexity of this dataset lies in differing sidechains and differing dihedral angles. AA has been used to benchmark several ML models \citep{bartok_machine_2017, de_comparing_2016, artrith_efficient_2017} and clustering techniques. \citep{de_mapping_2017}

\subsection{\label{sec:OE}OE: 64\,k opto-electronically active molecules}
This dataset, referred to as OE, consists of 64,710 large organic molecules with up to 174 atoms extracted from organic crystals in the Cambridge Structural Database (CSD). \citep{groom_cambridge_2016} Schober \textit{et al.} have screened the CSD for monomolecular organic crystals with the objective to identify organic semiconductors with high charge carrier mobility. \cite{Schober/etal:2016,Kunkel/etal_a:2019,Kunkel/etal_b:2019} For this study, we extracted molecules from the crystals and relaxed them in vacuum with the aforementioned computational parameters. The OE dataset is not yet publicly available. OE offers the largest chemical diversity among the sets in this work both in terms of size as well as number of different elements (Fig.~\ref{fig:statistics}). It contains the 16 different element types H, Li, B, C, N, O, F, Si, P, S, Cl, As, Se, Br, Te and~I. The electronic structures are more complex than in QM9 and AA, containing, e.g. large conjugated systems and unusual functional groups.

 \begin{figure*}[t!] 
\centerline{\includegraphics[width=16.5cm]{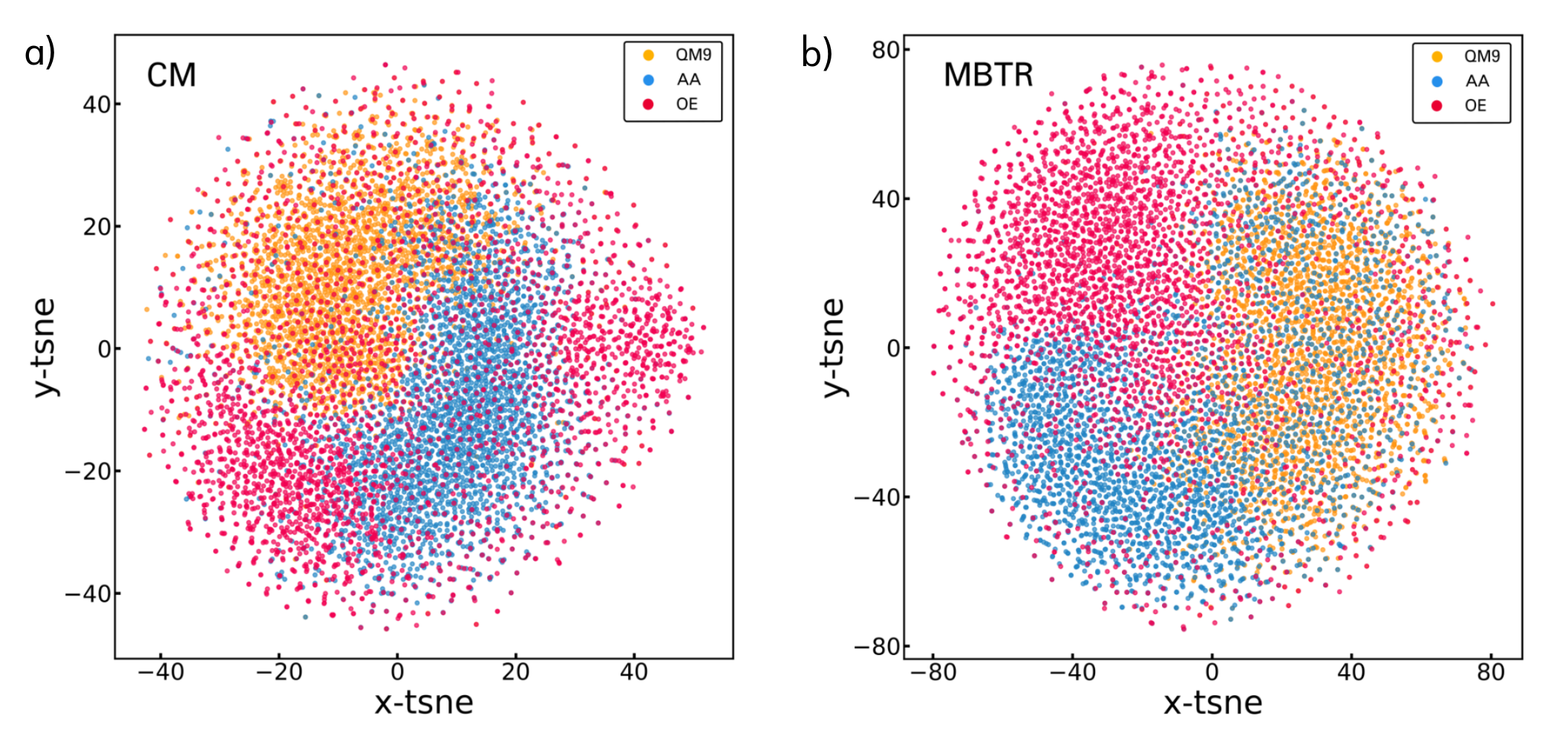}}
\caption{\label{fig:tsne}$t$-SNE analysis of the three datasets, where molecules are represented by a) CM and b) MBTR. OE molecules are widely spread out in both dimensions, while AA and QM9 molecules form their own groups. The t-SNE algorithm was run on 9,000 randomly sampled molecules, i.e. 3,000 molecules from each dataset.}
\end{figure*}

\subsection{\label{sec:datasets_comp}Comparison of datasets}
Figure~\ref{fig:statistics} illustrates the chemical diversity present in our datasets. The molecular size distribution of the three datasets in Fig.~\ref{fig:statistics} shows that QM9 and AA both contain molecules of similar sizes. The QM9 distribution exhibits a distinct peak at around 18 atoms, whereas AA has a bimodal distribution centered around 27 and 35 atoms. Conversely, the size distribution of OE is much broader, extending to molecules with as many as 174 atoms. In terms of element diversity, all datasets overlap on the 4 elements of organic chemistry: H, C, O, and N, as illustrated in Fig.~\ref{fig:statistics}. QM9 contains only F in addition, whereas AA branches out into common cations. OE offers the largest element diversity, including common semiconductor elements.

To compare our target property across the different datasets, we show distributions of the DFT pre-computed HOMO energies for each dataset in Figure \ref{fig:homo}. HOMO energies in QM9 and OE are centered around -6 eV, with a standard deviation of 0.5 eV and 0.6 eV, respectively. For AA, only a fraction of the HOMO energies are centered around -6 eV, while most HOMO energies are distributed over a wider range between -19.6 eV and -8 eV. The HOMO energies around -6 eV correspond to amino acids and dipeptides in bare organic configurations (free of cations). The HOMO energies between -19 and -8 eV correspond to amino acids and dipeptides with one of the six cations Ca$^{2+}$, Ba$^{2+}$, Sr$^{2+}$, Cd$^{2+}$, Pb$^{2+}$ or Hg$^{2+}$. The metal ions in AA shift the HOMO energies of the amino acids towards lower values.

Judging by the distributions of molecular size and element types in Fig. 2 and by the distributions of HOMO energies in Fig. 3, we expect QM9 to be learned relatively easily. QM9 is clustered both in chemical space (in terms of molecular size and element types) and in target space (in terms of HOMO energies). We therefore expect KRR to benefit from mapping similar input structures to similar target values. For OE, the target energy distribution is similar to QM9, but the molecular structures are widely spread through chemical space.  Similarity is therefore present only in one space, while the other is diverse. It might be challening for KRR to map from the diverse chemical space of OE to its confined target space and we expect learning to be slow. The AA set is spread out both in chemical space and target space. As it turns out, this will not be a problem for learning.

\begin{figure*}[t!] 
\centerline{\includegraphics[width=17.2cm]{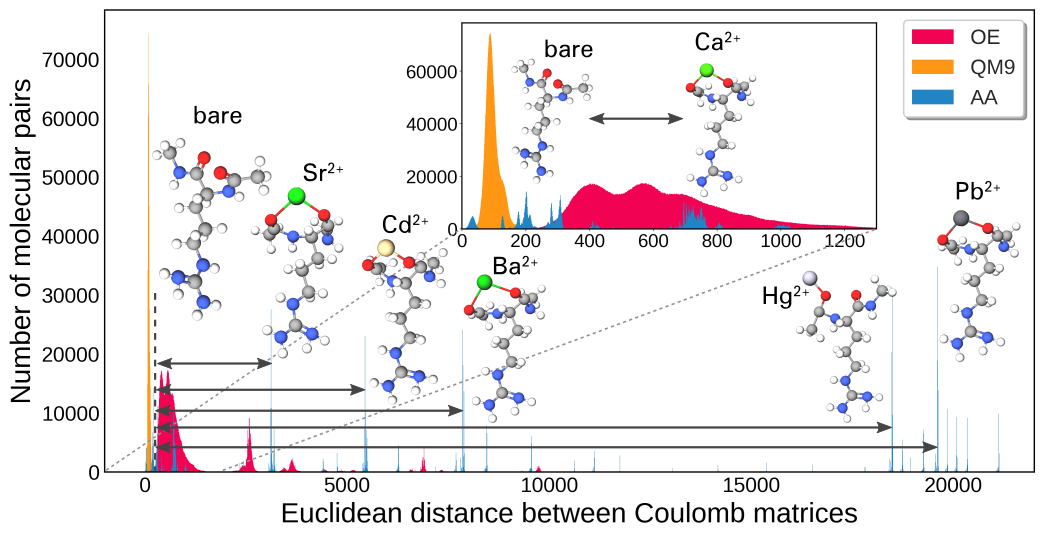}}
\caption{\label{fig:eucl_dist}Distributions of pairwise Euclidean distances computed for 3,000 randomly chosen molecules within each dataset. Molecules are represented by the CM. Molecular distances within QM9 are centered around small values, while OE distances are distributed evenly over a larger and wider range. The AA distribution consists of separate clusters, where each cluster corresponds to distances between amino acids with different metal ions attached to them. The inset shows an enlarged view of distances between 0 and 1300, where most of the distances within QM9 and OE are distributed. The two molecules are taken from the AA dataset and indicate the distance between bare structures and structures with a Ca$^{2+}$ ion attached.}
\end{figure*}

In Fig.~\ref{fig:tsne} the three datasets are visualized with the t-distributed stochastic neighbor embedding (t-SNE) dimensionality reduction technique for both the their CM and MBTR molecular representations. \citep{maaten_visualizing_2008} From each dataset, we randomly picked 3,000 molecules, which are mapped to a two dimensional space by t-SNE. Based on the pairwise similarities between descriptors, we can identify patterns within the datasets. In Fig.~\ref{fig:tsne}a), we can see that the AA and QM9 sets form distinct clusters and that the two clusters (orange and blue) have almost no overlap in the CM representation. The MBTR  t-SNE analysis produces the same result, as indicated by Fig.~\ref{fig:tsne}b) (cluster orientation is arbitrary). This implies that the molecules in QM9 are very different from those in AA and that we increase the chemical diversity of our study by including the AA dataset. Fig.~\ref{fig:tsne} further illustrates that the OE set is maximally diverse in itself. The corresponding red point cloud is evenly spread over the whole figure. In contrast to the QM9 and AA sets, OE offers a variety of rigid backbones (also denoted scaffolds) to which different functional groups are attached. Chemical diversity in part arises due to the rich combinatorial space emerging from scaffold-functional group pairings. For example, OE contains molecules with conjugated and aromatic backbones and electron withdrawing and donating functional groups that are of technological relevance and completely absent from QM9 and AA \citep{Kunkel/etal_a:2019,Kunkel/etal_b:2019}.

\vspace{-0.75cm}
To quantify the similarity between molecules in our three datasets, we computed Euclidean distances for molecular pairs -- represented by the CM -- for 3,000 randomly chosen molecules for each dataset, as shown in Fig. \ref{fig:eucl_dist}. Molecular distances within QM9 are small, indicating great similarity among QM9 molecules. The OE distances are distributed evenly over a larger and wider range of distances, indicating high dissimilarity among OE molecules. The AA distance distribution has several separated clusters up to very large distances. The distances are separated and ordered by the atomic number of the cations, where each cluster corresponds to amino acids with different metal ions attached. The first cluster includes distances from bare amino acids to amino acids with Ca$^{2+}$, followed by clusters including distances to amino acids with Sr$^{2+}$, Cd$^{2+}$, Ba$^{2+}$, Hg$^{2+}$ and Pb$^{2+}$. Structural dissimilarity in the AA dataset mainly arises due to amino acids with different metal ions, while amino acids within the same cluster are highly similar to each other. \\
\newline

\vspace{-0.1cm}
\section{\label{sec:representation}Molecular representation}
For the ML model to make accurate predictions, it is important to represent the molecules for the machine in an appropriate way. \citep{rupp_machine_2015, bartok_on_2013, lilienfeld_fourier_2015, huo_unified_2017} Cartesian (x,y,z)-coordinates, which are, for example, used for DFT calculations, are not applicable, since they are not invariant to translations, rotations, and reordering of atoms. In this work, we compare the performance of two different molecular representations. 

\vspace{-0.3cm}
\subsection{Coulomb matrix (CM)}
\begin{figure}[ht] 
    \centerline{\hspace*{0.2cm}\includegraphics[width=8.4cm]{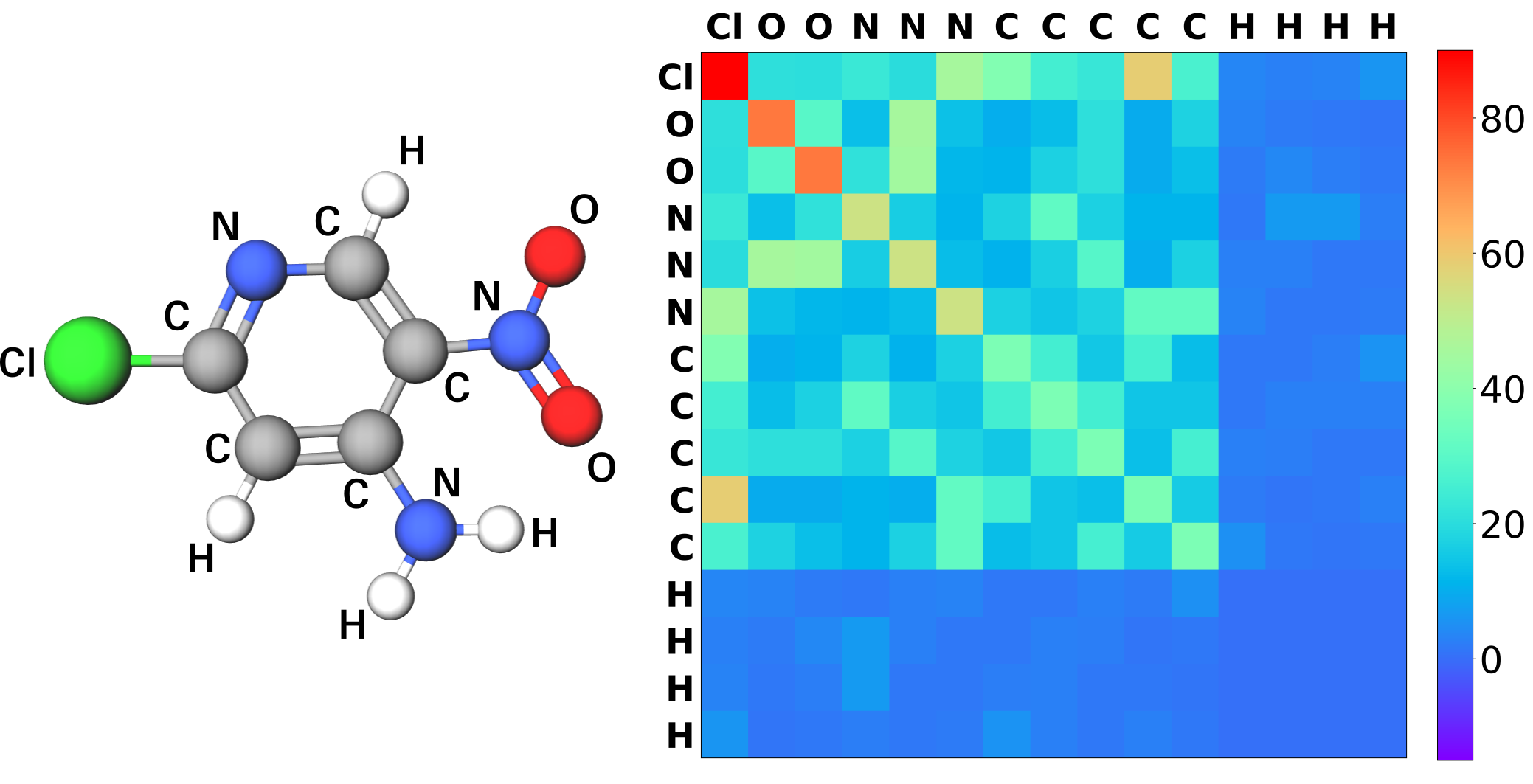}}
    \caption{\label{fig:cm}Coulomb matrix representation (right) of the molecule 2-chloro-5-nitropyridin-4-amine (left) taken from OE.}
\end{figure}

In the CM formalism, \citep{rupp_fast_2012} each molecule is represented by a matrix {C},
\begin{equation}
  C_{ij} = 
  \begin{cases}
    0.5Z_i^{2.4} & \text{if $i=j$}\\
    \frac{Z_i Z_j}{\lVert \mathbf{R_i}-\mathbf{R_j} \rVert } & \text{if $i \neq j$}\\
  \end{cases} .
\end{equation}
The CM encodes nuclear charges $Z_i$ and corresponding Cartesian coordinates $R_i$, with off-diagonal elements representing Coulomb repulsion between atom pairs and diagonal elements encoding a polynomial fit of free-atom energies to~$Z$. An example of the CM for a molecule of OE is shown in Fig.~\ref{fig:cm}. To enforce permutational invariance, we simultaneously sort rows and columns of all CMs with respect to their $\ell^2$-norm.

\subsection{Many-body tensor representation (MBTR)}
 
\begin{figure}[h!]
    \centerline{\includegraphics[width=7.5cm]{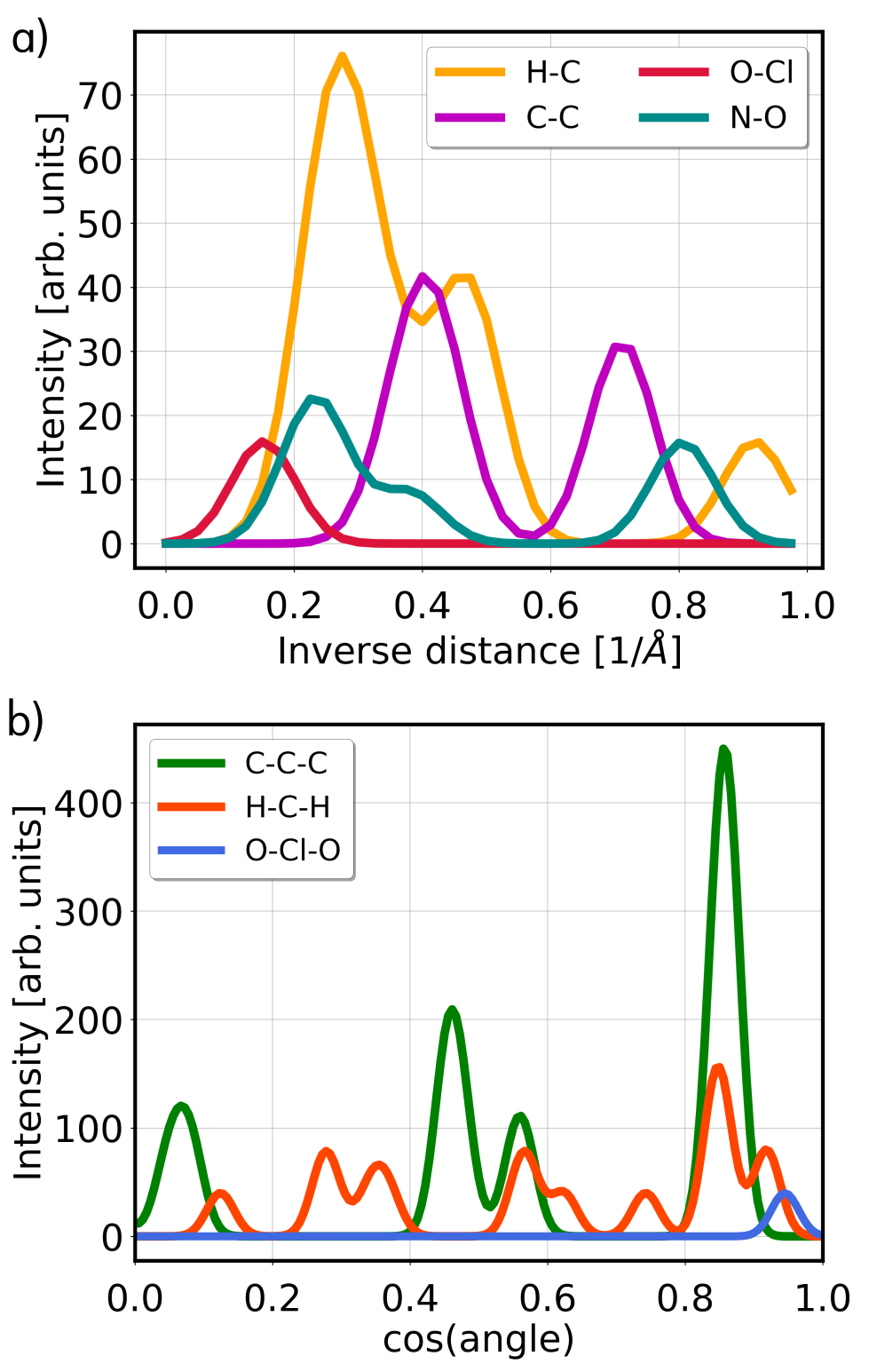}}
\caption{\label{fig:mbtr}Many-body tensor representation of the molecule from Fig.~\ref{fig:cm} (2-chloro-5-nitropyridin-4-amine). a) Inverse distance distribution of selected two-body terms with optimized broadening parameter $\rho_2$. b) Angle distribution of selected three-body terms with optimized broadening parameter $\rho_3$.}
 \end{figure}

The MBTR \citep{huo_unified_2017} can be viewed as a many-body expansion of the Bag of Bonds (BoB) representation, \citep{hansen_machine_2015} which in turn is based on elements of the CM. 
One-body terms of the MBTR describe the atom types that are present in the molecule. 
Two-body terms encode inverse distances between all pairs of atoms (bonded and non-bonded), separately for each combination of atom types. The inverse distances are sorted by increasing order and broadened into a continuous Gaussian distribution of inverse distances as shown in Fig.~\ref{fig:mbtr}a). Three-body terms encode angle distributions for any triplets of atoms present in the molecule, as shown in Fig.~\ref{fig:mbtr}b). Each $N$-body term has a broadening parameter (in total $\rho_1$, $\rho_2$ and $\rho_3$) that controls the smearing of atom type distribution, inverse distance distribution and angle distribution, respectively, and need to be fine-tuned for optimal KRR performance. We use the \emph{DScribe} package \cite{dscribe} to compute the MBTR and the \emph{qmmlpack} package to refine the MBTR hyperparameters for small training set sizes of up to 4k molecules. Exponential weighting was employed for the computation of inverse distance and angle terms. Here, we apply only two-body and three-body terms in the MBTR, since we found that the inclusion of one-body terms does not improve the performance, but increases computational time (we refer to Appendix \ref{sec:kernels} for details).

\section{\label{sec:method}Machine learning method}
\subsection{Kernel ridge regression} 

We employ kernel ridge regression \citep{hastie_elements_2009} (KRR) to model the relationship between molecular structures and HOMO energies. In KRR, training samples are mapped into a high-dimensional space using nonlinear mapping, and the structure-HOMO relationship is learned in the high-dimensional space. This learning procedure is conducted implicitly by defining a kernel function, which measures the similarity of training samples in the high-dimensional space by employing a kernel function. In this work, we use two different kernel functions. The first kernel is the Gaussian kernel

\begin{equation}
k_{\textnormal{Gaussian}}(\boldsymbol{x},\boldsymbol{x}')=e^{-\frac{||{\boldsymbol{x}-\boldsymbol{x}'}||_2^2}{2\sigma^2}},
    \label{eq:gaussian}
\end{equation}
where
\begin{equation}
 ||{\boldsymbol{x}-\boldsymbol{x}'}||_2= \sqrt{\sum_i |\boldsymbol{x_i} - \boldsymbol{x_i}'|^2}
 \label{eq:similarity-2}
\end{equation}
is the Euclidean distance and $\boldsymbol{x}$, $\boldsymbol{x}'$ are two training molecules represented by either the CM or the MBTR. We note that the Euclidean distance distribution between molecular pairs shown in Fig. \ref{fig:eucl_dist} gives us a direct insight into the learning process of a Gaussian kernel function. The second kernel is the Laplacian kernel
\begin{equation}
    k_{\textnormal{Laplacian}}(\boldsymbol{x},\boldsymbol{x}')=e^{-\frac{||{\boldsymbol{x}-\boldsymbol{x}'}||_1}{\sigma}}.
    \label{eq:laplacian}
\end{equation}
which uses the 1-norm as similarity measure,
\begin{equation}
 ||{\boldsymbol{x}-\boldsymbol{x}'}||_1= \sum_i |\boldsymbol{x_i} - \boldsymbol{x_i}'|.
 \label{eq:similarity-1}
\end{equation}
In Eqs.~\eqref{eq:gaussian} and~\eqref{eq:laplacian}, $\sigma$ is the kernel width. 

In the KRR training phase with $N$ training molecules, the goal is to find a vector $\bm{\alpha} \in \mathbb{R}^N$ of regression weights $\alpha_i$ that solves the minimization problem
\begin{equation}
 \underset{\alpha}{\textnormal{min}} \sum_{i=1}^N (E^{\textnormal{pred}} (\boldsymbol{x}_i) - E^{\textnormal{ref}}_i)^2 + \lambda \bm{\alpha}^T \mathbf{K} \bm{\alpha},
 \label{eq:krr_minimization}
\end{equation}
where the analytic solution for $\bm{\alpha}$ is given by 
\begin{equation}
\bm{\alpha} = (\mathbf{K} + \lambda \mathbf{I})^{-1} \mathbf{E}^{\textnormal{ref}}.
\end{equation}

The matrix $\mathbf{K}$ is the kernel matrix, whose elements represent inner products between training samples in the high-dimensional space, calculated as kernel evaluations $K_{i,j}:=k(\boldsymbol{x_i}, \boldsymbol{x_j}$). The scalar $\lambda$ is the regularization parameter, which penalizes complex models with large regression weights over simpler models with small regression weights. The expression $E^{\textnormal{ref}}$ denotes the reference HOMO energy computed by DFT and $E^{\textnormal{pred}}$ is the predicted HOMO energy, which is obtained as sum over weighted kernel functions
\begin{equation}
E^{\textnormal{pred}}(\boldsymbol{x})=\sum^{N}_{i=1} \alpha_i k(\boldsymbol{x},\boldsymbol{x}_i).
 \label{eq:predict}
\end{equation}
The sum runs over all molecules $\boldsymbol{x}_i$ in the training set with their corresponding regression weights  $\alpha_i$. After training, predictions are made for test molecules that were not used to train the model, employing eq.~\eqref{eq:predict} to predict out-of-sample molecules and estimate performance of the model. No further scaling or normalization of the data was done to preserve the meaning of the HOMO energies. 

\subsection{KRR training, cross-validation and error evaluation}

For each dataset QM9, AA and OE, we randomly selected a subset of 32\,k molecules for training and a further 10\,k molecules for out-of-sample testing. In order to obtain statistically meaningful results for the training and testing performance of KRR, we repeated the random selection of training set and test set nine more times for each dataset after the data was reshuffled. As a result, we acquired 10 different training sets of 32k molecules and 10 different test sets of 10k molecules for each dataset. 

\begin{figure*}[t!] 
    \centerline{\includegraphics[width=18.1cm]{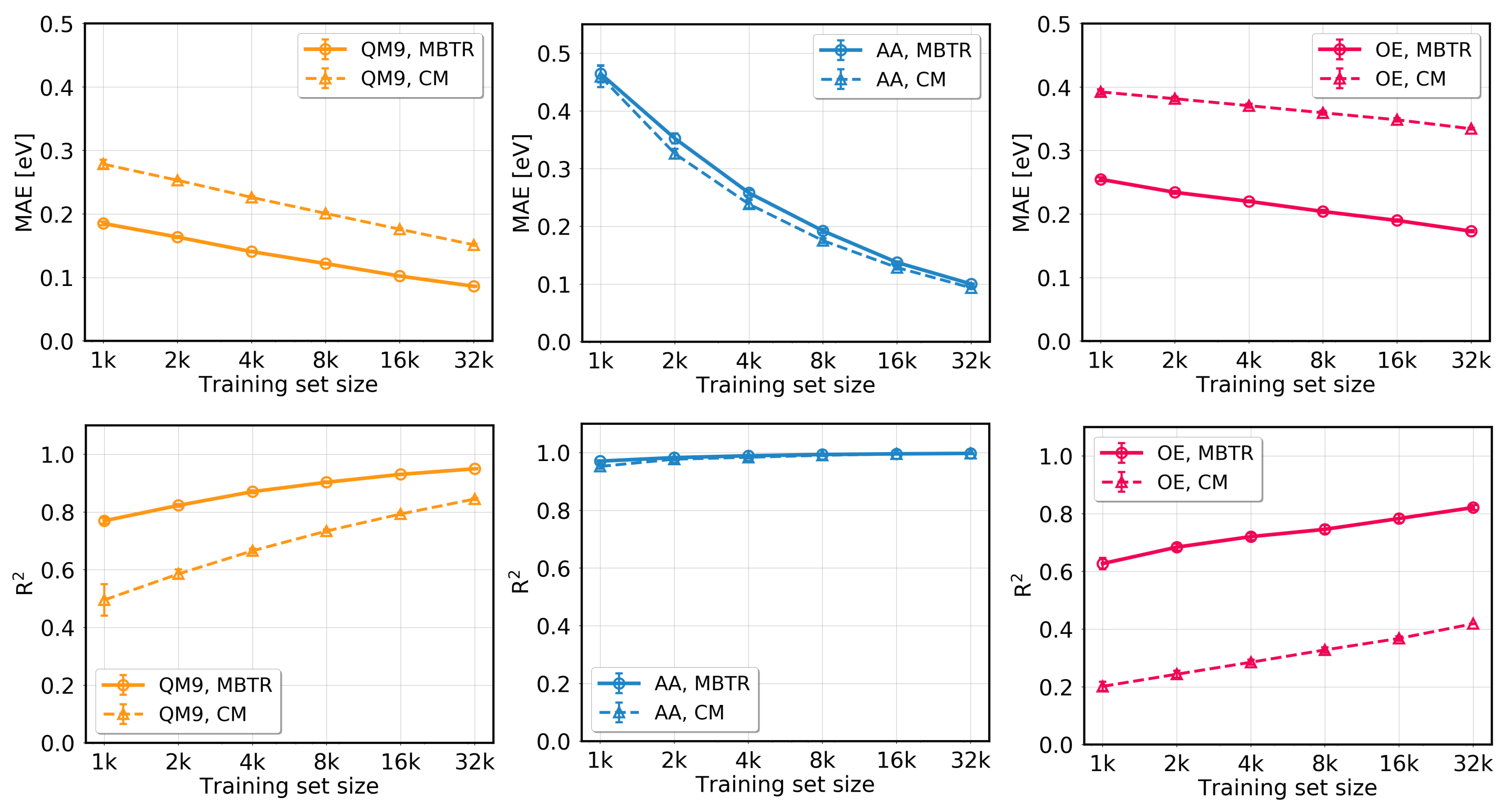}}
\caption{\label{fig:mae_r2}Mean absolute errors (MAE) (upper panel) and squared correlation coefficient $R^2$ (lower panel) for out-of-sample predictions of the HOMO energy as a function of training set size for QM9 (orange), AA (blue) and OE (red). For each dataset, performances of MBTR (filled lines) and CM (dashed lines) are compared to each other. The Laplacian kernel was used in combination with the CM and the Gaussian kernel was used in combination with the MBTR. The MAE datapoint and its error bar correspond to the average value and its standard deviation computed over 10 instances of training, cross validating and out-of-sample testing.}
\end{figure*}

From each training set of 32k, we randomly drew 6 different subsets of sizes 1k, 2k, 4k, 8k and 16k, where smaller sets were always subsets of larger ones. For each of these subsets, we trained and cross-validated a KRR model, as described in the following paragraph. We then evaluated the KRR model on the corresponding test set of 10k by predicting HOMO energies for 10k out-of-sample molecules and by computing the MAE $\frac{1}{n}\sum_{i=1}^{n}|E^{\textnormal{pred}}_i-E^{\textnormal{ref}}_i|$ between predictions and DFT reference energies. For each dataset and for each training set size, we computed the average MAE value and its standard deviation across the 10 randomly drawn training and test sets. We then plotted all of these average MAEs as a function of training set size and as a result, attain one learning curve for each dataset, as shown in Fig. \ref{fig:mae_r2} and Fig. \ref{fig:mae}. We note that for the two molecular representations, CM and MBTR, identical training and test sets were used. 

In the scope of this study, there are 3 types of hyperparameters: 
\begin{enumerate}[label=(\roman*)] \vspace{0.2cm}
\item the kernel function (Gaussian type or Laplacian type)
\item  MBTR hyperparameters (broadening values $\rho_1$, $\rho_2$ and $\rho_3$)  \vspace{-0.2cm}
\item KRR hyperparameters (kernel width $\sigma$ and regularization parameter $\lambda$) \end{enumerate}
The large number of possible values for the hyperparameters leads to a wide range of possible KRR models. The choice of kernel function depends on the molecular descriptor. The Gaussian kernel performs best on the MBTR, whose $N$-body terms themselves consist of Gaussian distributions. The Laplacian kernel, on the other hand, can better model piecewise smooth functions, such as discontinuities of the sorted CM. Therefore, we chose the Laplacian kernel for the CM and the Gaussian kernel for the MBTR. We provide learning curves in Appendix \ref{sec:kernels} to prove that we picked the optimal kernel for each descriptor.

The MBTR and KRR hyperparameters were simultaneously optimized in a cross-validated grid search for each of the three smaller training sets of 1k, 2k and 4k. In particular, MBTR hyperparameters $\rho_1$, $\rho_2$ and $\rho_3$ were varied on a grid of 10 points between $10^{-4}$ and $10^{-1}$ and KRR hyperparameters $\sigma$ and $\lambda$ were varied on a grid of 12 points between $10^{-12}$ and $10^{-1}$. The aim was to find the combination of best KRR and MBTR parameters. For more details on the cross-validated grid search we refer to Appendix \ref{sec:cv}. The optimized model was then evaluated on 10k out-of-sample molecules from the test set. Finally, the MAE on the test set was reported as a data point on the learning curve for a given training set size.

We found that the values of the optimized MBTR hyperparameters do not change throughout the three small training sets. We therefore fixed the optimized MBTR hyperparameters for the larger larger training sets of 8k, 16k and 32k and optimized only the KRR hyperparameters (on a grid from $10^{-12}$ to $10^{-1}$). 

Another measurent for the performance of a ML model is the $R^2$ coefficient, which describes the proportion of variability in a dataset that can be explained by the model. Although we do not use the R$^2$ coefficient to optimize KRR hyperparameters, we employ this metric to interpret how well our models fit the data. \\


\vspace{-0.2cm}
\begin{center}
\begin{table*}[ht!]
\setlength{\tabcolsep}{4pt}
\begin{tabular}{ccccccc}
\multicolumn{4}{@{}l}{}\\
\hline
\hline
 &  \multicolumn{2}{c}{\textbf{MAE [eV]}} & \multicolumn{2}{c}{\textbf{RMSE [eV]}} &  \multicolumn{2}{c}{\textbf{R$^2$}} \\
\hline
\textbf{Dataset} & \textbf{MBTR}  & \textbf{CM} & \textbf{MBTR}  & \textbf{CM} & \textbf{MBTR}  & \textbf{CM} \\
\hline
QM9 & 0.086 $\pm{~0.001}$ & 0.151 $\pm{~0.001}$ & 0.118 $\pm{~0.002}$ & 0.207 $\pm{~0.003}$ & 0.950 $\pm{~0.002}$ & 0.845 $\pm{~0.004}$\\
AA  & 0.100 $\pm{~0.001}$ & 0.094 $\pm{~0.002}$ & 0.201 $\pm{~0.006}$ & 0.194 $\pm{~0.007}$ & 0.997 $\pm{~0.001}$ & 0.998 $\pm{~0.001}$  \\
OE  & 0.173 $\pm{~0.002}$ & 0.336 $\pm{~0.003}$ & 0.239 $\pm{~0.006}$ & 0.435 $\pm{~0.005}$ & 0.821 $\pm{~0.009}$ & 0.413 $\pm{~0.005}$ \\
\hline
\hline
\end{tabular}
\captionof{table}{\label{table:results}Mean absolute errors (MAEs), root mean square errors (RMSEs) and R$^2$ coefficients for KRR predictions of molecules from the QM9, AA and OE datasets. Results are shown for the MBTR and CM representations. Errors and R$^2$ coefficients are measured on test sets of 10\,k randomly selected out-of-sample molecules of each dataset, while 32\,k molecules were used for training. We report errors and $R^2$ coefficients as the average over 10 repetitions, accompanied by the standard deviation of the mean. The corresponding reference-versus-predicted scatter plots are shown in Fig.~\ref{fig:scatter} for the first run out of 10 repetitions.} 
\end{table*}
\end{center}

\vspace{-1cm}
\section{\label{sec:results}Results}
Upper panels in Fig.~\ref{fig:mae_r2} show out-of-sample MAEs as a function of training set size ("learning curves'') for the different datasets and for CM and MBTR as descriptors. As expected, the MAE decreases for all datasets with increasing training set size (see also Fig.~\ref{fig:mae}). The best MAEs for a training set size of 32k molecules are presented in Table~\ref{table:results}. The lowest MAE is achieved for QM9, closely followed by AA. In constrast, the MAEs for OE are approximately twice as high. The learning rate (slope of the MAE curves) is highest for AA and lowest for OE, and is independent of the descriptor. The MBTR performs significantly better than the CM for QM9 and OE, while for AA, the learning curves of the two descriptors are the same within statistical errors.

\begin{figure*}[ht!] 
    \centerline{\includegraphics[width=17.6cm]{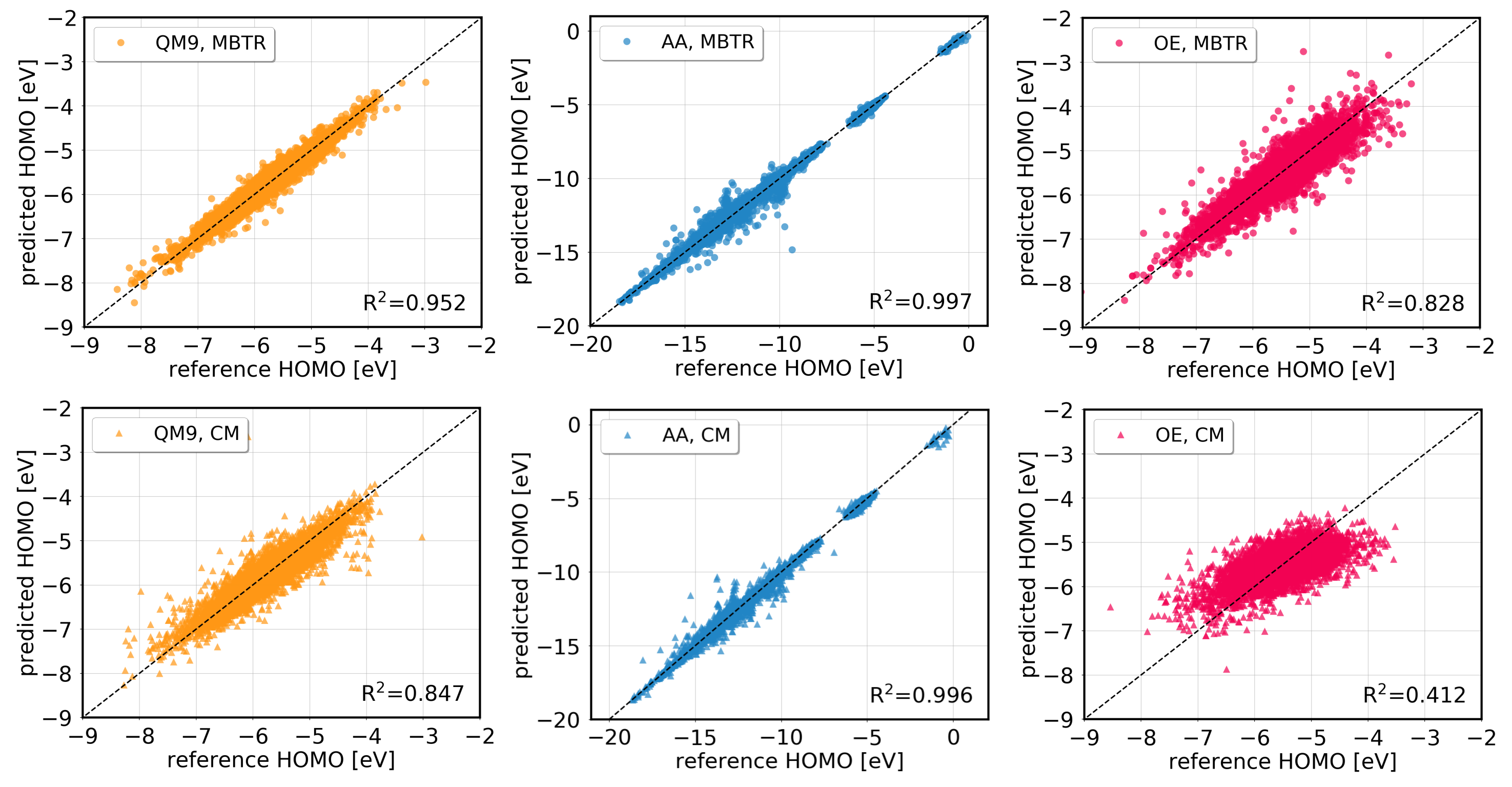}}
\caption{\label{fig:scatter}Scatter plots of out-of-sample predictions of the HOMO energy on 10~k molecules of QM9 (orange), AA (blue) and OE (red). We show reference HOMO energies pre-calculated with DFT versus model predictions for the first run out of 10 repetitions. The training set size is 32\,k. The upper panel shows results obtained with the MBTR and the lower panel shows results obtained with the CM. The Laplacian kernel was used in combination with the CM and the Gaussian kernel was used in combination with the MBTR.}
\end{figure*}

The lower panels of Fig.~\ref{fig:mae_r2} show the squared correlation coefficient $R^2$ as a function of training set size. $R^2$ increases systematically with training set size, but its rate varies across the three datasets. For AA it is close to 1 already for small training set sizes, whereas for QM9, $R^2$ starts off low and then approaches 1 for 32\,k. For the OE dataset, $R^2$ steadily increases and reaches a value of 0.81 at 32\,k.

\vspace{-0.1cm}
Correlating $R^2$ and MAE, we observe that model fitting for OE is consistently poor (low $R^2$) and the prediction errors are consistently high (high MAEs). The CM appears to be less suitable for this prediction task.

\vspace{-0.1cm}
Fig.~\ref{fig:scatter} presents scatter plots for a fixed training set size of 32\,k, which describes how well the KRR predictions correlate with the reference values in the test set. We observe the best correlation (i.e. predictive power) for AA and the worst for OE. The MBTR appears to have higher predictive power than CM for QM9 and OE, whereas MBTR and CM perform similarly for AA.

\vspace{-0.1cm}
Fig.~\ref{fig:mae} summarizes the learning curves for the three datasets. The top panel shows the MBTR and the bottom panel the CM results. 
The MBTR-based KRR models generally produce faster learning rates and lower MAEs. The AA learning rate is particularly fast, and eventually leads to MAEs of 0.10~eV for MBTR and 0.09 eV for CM, which are comparable to those of QM9. The prediction quality for the OE dataset is notably worse in relation to the other datasets.

\begin{figure*}[] 
    \centerline{\includegraphics[width=17.6cm]{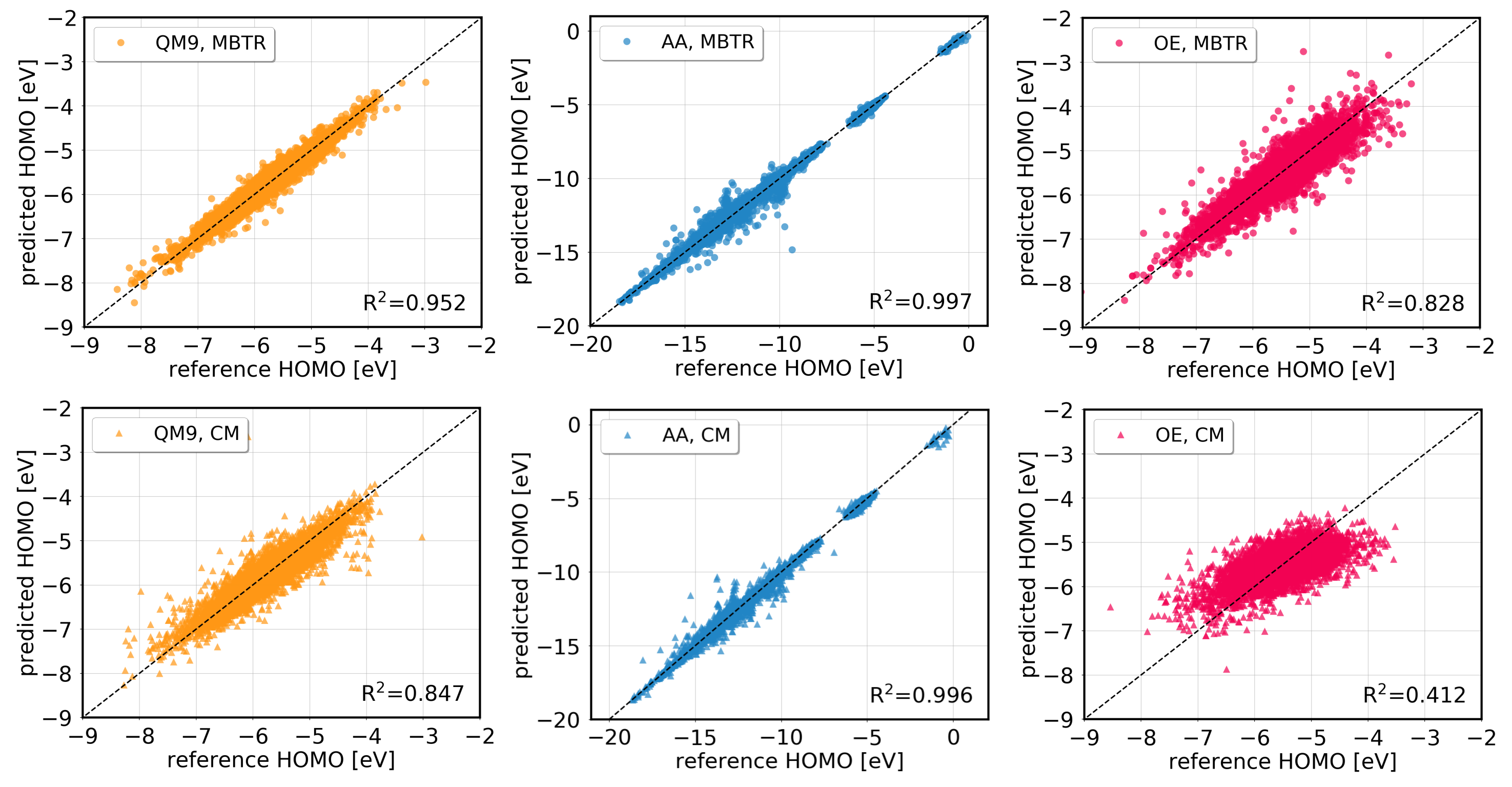}}
\caption{\label{fig:scatter}Scatter plots of out-of-sample predictions of the HOMO energy on 10~k molecules of QM9 (orange), AA (blue) and OE (red). We show reference HOMO energies pre-calculated with DFT versus model predictions for the first run out of 10 repetitions. The training set size is 32\,k. The upper panel shows results obtained with the MBTR and the lower panel shows results obtained with the CM. The Laplacian kernel was used in combination with the CM and the Gaussian kernel was used in combination with the MBTR.}
\end{figure*}

\begin{figure}[ht!]\hspace{-0.6cm}
      \centerline{\includegraphics[width=8cm]{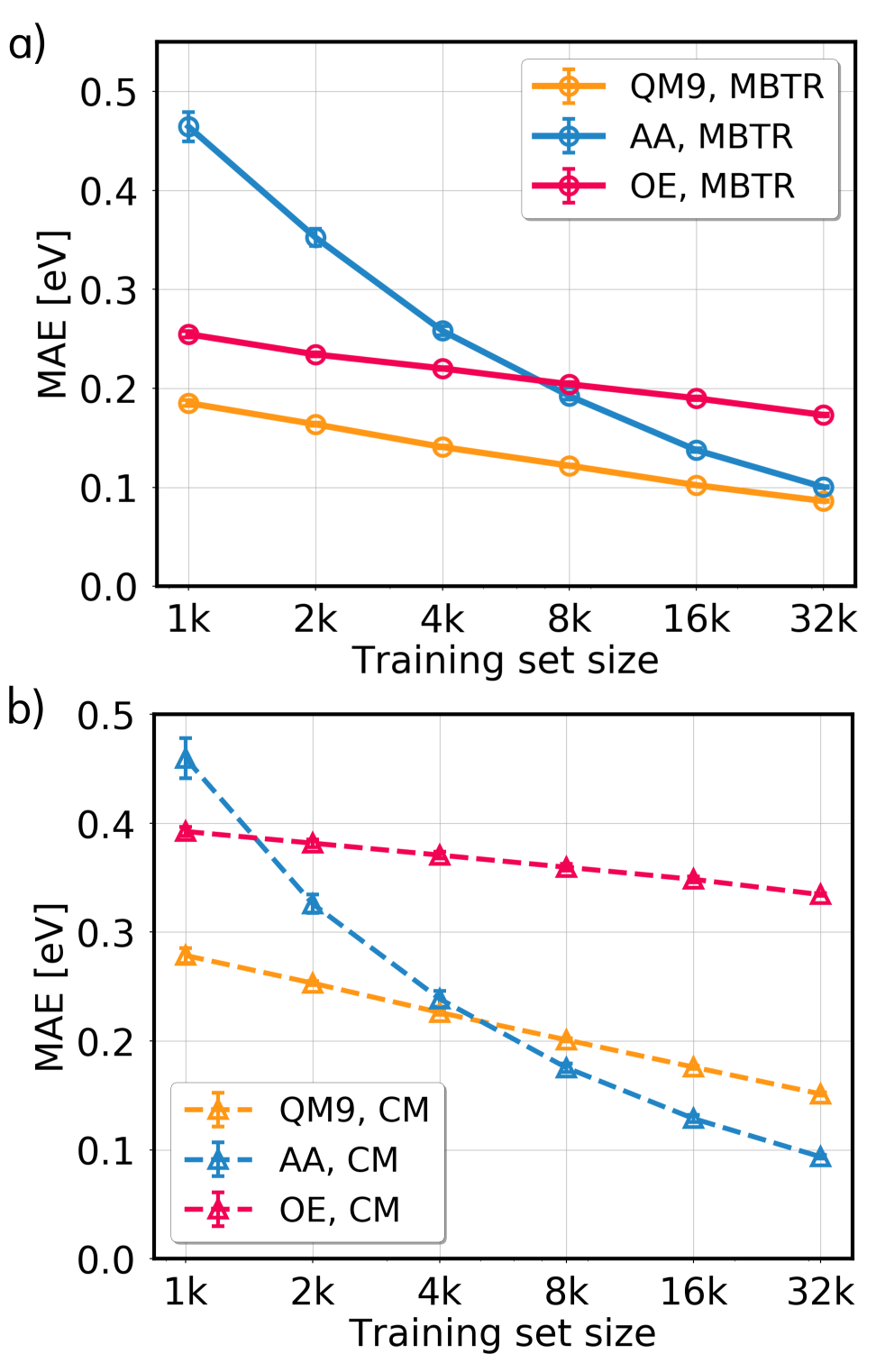}}
\caption{\label{fig:mae}Comparison of learning curves for QM9 (orange), AA (blue) and OE (red). KRR models are based on a) Gaussian kernel in combination with MBTR and b) Laplacian kernel in combination with CM. The error decay is shown on a logarithmic scale. The MAE datapoint and its error bar correspond to the average value and its standard deviation computed over 10 training instances.}
 \end{figure}

\section{Discussion}
\label{sec:discussion}
\subsection{Dependence of KRR performance on dataset diversity}

The learning success of KRR depends on the structural complexity of individual molecules (e.g. number of atoms, atom types, backbone types etc.) as well as on the diversity and redundancy within a dataset. Redundancy usually means that certain structural features occur frequently in a dataset, i.e. many data points are similar to each other. Diversity implies the opposite: Few instances in a dataset are similar to the rest. Redundant datasets are learned well with ML, even when trained on small portions. Diverse datasets can pose a problem for ML, even when applied to large data.
 
The differences in the learning curves we observe in Figs.~\ref{fig:mae_r2} and \ref{fig:mae} reflect the chemical differences in the three datasets employed in our study. QM9 is greatly redundant and includes molecules with simple structures. Therefore, it can be learned well even on small training set sizes, where redundancy is low.
 
The AA dataset has inbuilt redundancy, but also includes many different metal cations. For small training set sizes, where not enough similar structures per metal cation are present, the error is high. This situation then improves quickly with increasing traing set size. As a result, the learning rate of AA is faster than of QM9.

OE is highly diverse and includes molecules with complex structures. It has similar chemical element range as AA with 16 different element types, but features larger molecules and more structural diversity, as discussed in Section~\ref{sec:datasets_comp}. The diversity explains the high errors throughout all training set sizes and the slower learning rate. The t-SNE analysis in Fig.~\ref{fig:tsne} confirms that OE has little overlap with QM9, regardless of the molecular representation. OE is the most diverse of the three datasets, since the corresponding point cloud does not cluster in a particular region and instead fills the whole space of Fig.~\ref{fig:tsne}.

The KRR learning rate is slowest for OE, due to the aforementioned chemical and structural complexity in this dataset. At a training set size of 32\,k, the MAE is still twice as high as for QM9 and AA. With 0.173\,eV, the MAE for OE is still too high for the spectroscopic applications we intend, which typically require errors below 0.1\,eV. Advanced machine learning methodology and more sophisticated materials descriptors both help to reduce overall prediction errors \citep{montavon_learning_2012, faber_prediction_2017, faber_alchemical_2018, schutt_schnet_2018, pyzer_learning_2015} and we may test alternative approaches in further work.

Lastly, our study of three different datasets illustrates that MAEs are not transferable across datasets, even if they are evaluated for the same machine learning method and the same descriptor. If we had based our predictive power expectations for the OE set on the KRR QM9 performance, we would have been disappointed to find much larger errors in reality. It is therefore paramount to further investigate the performance of machine learning methods across chemical space.

\vspace{-0.5cm}
\subsection{Dependence of KRR performance on molecular descriptor}

Next, we discuss the relative performance for the CM and MBTR molecular descriptors. Overall, MBTR outperforms the CM across the datasets, which is in line with previous findings. \citep{huang_communication_2016, faber_prediction_2017, pereira_machine_2017, collins_constant_2018} This is reasonable in light of the higher information content about atom types, their bond lenghts and angles encoded in the MBTR when compared to the CM matrix. The CM and the MBTR exhibit the same performance only for the AA dataset of molecular conformers, where complexity is dominated by the torsional angles, and bonding patterns are similar. This result is partly explained by the exclusion of torsional angle information into MBTR (four-body terms), while consistent chemical information in AA benefits the performance of the CM.

It is interesting that CM generally produces MAEs only twice as large as the MBTR with much smaller data structures and at a fraction of the computational cost. The CM representation is simple to compute, supplies benchmark results comparable to previous work and may prove a convenient tool for preliminary studies of large unknown datasets.

\subsection{Application of QM9 model}

Our results for the QM9 dataset allow us to compare our findings with previous studies. Given a KRR training set of HOMO energies for 32\,k molecules, we obtained MAE values of 0.086\,eV and 0.151\,eV with the MBTR and CM descriptors respectively.
In 2017, Faber \textit{et al.} \citep{faber_prediction_2017} reported KRR results on 118\,k QM9 molecules, using a molecular descriptor based on interatomic many-body expansions including bonding, angular and higher-order terms, \citep{huang_communication_2016} which is comparable to the MBTR used in this work. The HOMO energy was predicted with an out-of-sample MAE of 0.095\,eV, and the CM representation achieved an MAE of 0.133\,eV. Our QM9 results are in very good agreement with this study, even if our KRR training set is much smaller. These errors are relatively small and comparable to experimental and computational errors in HOMO determinaton, which typically range in between several tenth of eV. Errors in HOMO energy predictions with machine learning may be further reduced by developing customized deep learning neural network architectures, which have been reported to produce an MAE of 0.041 eV after training on 110~k QM9 molecules. \citep{schutt_schnet_2018}

\begin{figure}[ht!]
\centerline{\hspace*{-0.2cm}\includegraphics[width=9cm]{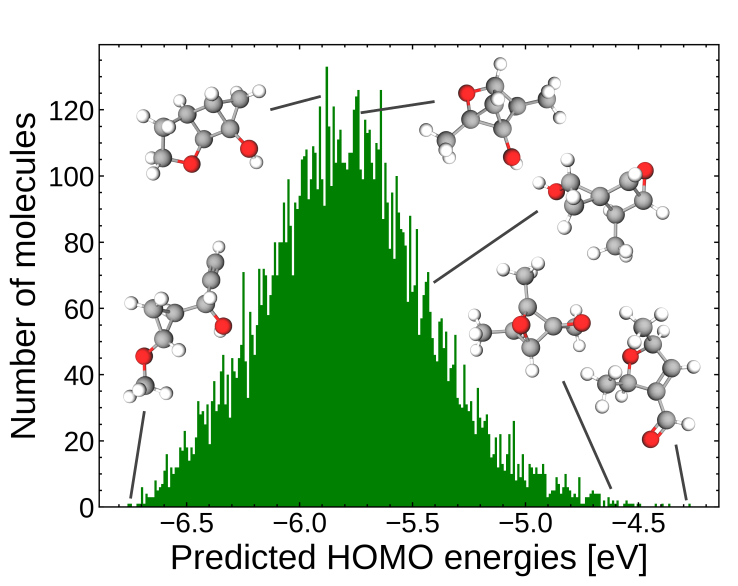}}
\caption{\label{fig:10k_diastereomers}HOMO energies predicted by our QM9-trained KRR model for a new dataset of 10k diastereomers, for which only molecular structures, but no pre-computed HOMO energies are available. Molecules that fall within a certain energy range can easily be identified and might be further assessed for potential applications, as illustrated for six example molecules.}
\end{figure}

To showcase the value of our trained KRR model, we apply it to a dataset of 10k diastereomers of parent C$_7$H$_{10}$O$_2$ isomers. 
This dataset contains molecular structures, but no HOMO energies. Computing the HOMO energies with DFT would take considerable effort and time. With our KRR model -- trained on 32k QM9 molecules represented by the MBTR -- we gain an immediate overview of the HOMO energies that occur in the dataset. A histogram of all predicted HOMO energies is shown in Fig. \ref{fig:10k_diastereomers}. We can see that they are uniformly distributed between -6.8 and -3.2 eV for all diastereomers. The energetic scan allows us to quickly detect molecules of interest in a large collection of compounds. Individual molecules can be easily identified, for instance, the molecule with lowest HOMO energy, molecules with highest HOMO energy or those molecules with average HOMO energies. Various molecules of interest, e.g., structures with HOMO energies in a particular region, could subsequently be further investigated with first-principle methods or experiments to determine their functionality for certain applications. In this fashion, fast energy predictions of our KRR model can be analyzed for structures with desired HOMO energy.

\section{\label{sec:conclusion}Conclusion}

In this study, we have trained and tested KRR models on three molecular datasets with different chemical composition to predict molecular HOMO energies. Our comparison between two different molecular descriptors, the CM and the MBTR, shows that the MBTR outperforms the CM on both OE and QM9 due to its ability to encode complex information from the molecular structures. For AA we could find no significant difference in performance between the two representations.

Our work demonstrates that the predictive performance of KRR inherently depends on the complexity of the dataset it is applied to, in addition to the training set size and descriptor. Rapidly decreasing learning curves and low MAEs are achieved for QM9, which is known as a standard benchmark set for ML in molecular chemistry, containing  pharmaceutically relevant compounds with rather simple bonding patterns. The same is true for AA which consists of a primitive and restricted collection of amino acids and peptides. The OE dataset, however, comprises large opto-electronic molecules with complicated electronic structures and unconventional functional groups, is much more difficult to learn. It yields almost flat learning curves and considerably higher MAEs. 
To further improve the predictive power for molecules of technological interest, such as the ones in the OE set, future work should focus on generating larger datasets, devising better descriptors or more sophisticated machine learning methods.

\begin{acknowledgments}
We greatfully acknowledge the CSC-IT Center for Science, Finland, and the Aalto Science-IT project for generous computational resources. This study has received funding from the European Union's Horizon 2020 research and innovation programme under grant agreement No 676580 with The Novel Materials Discovery (NOMAD) Laboratory, a European Center of Excellence and from the Magnus Ehrnrooth Foundation. This work was furthermore supported by the Academy of Finland through its Centres of Excellence Programme 2015-2017 under project number 284621.

\end{acknowledgments}

\appendix
\section{\label{sec:chem_names}Chemical names of molecules}
\begin{figure}[!htb] 
\centering
    \includegraphics[width=7.7cm]{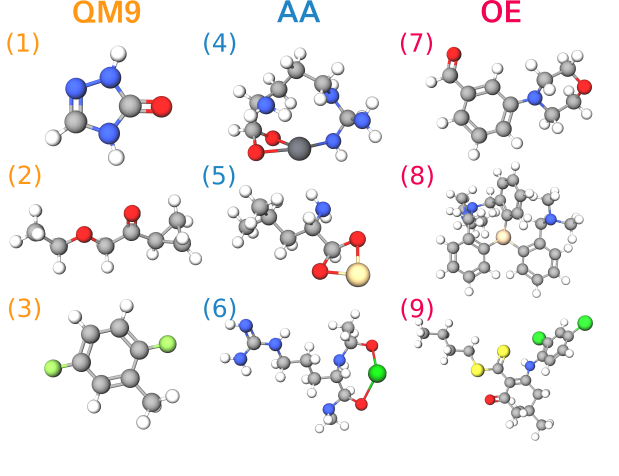}
    \caption{\label{fig:molecules_app} Example molecules taken from the three datasets used in this work. Chemical names are (1) 1H-1,2,4-Triazol-3-ol, (2) 1-Cyclopropyl-2-ethoxyethanone, (3) 2,5-Difluorotoluene, (4) Arginine (uncapped), (5) Leucine (uncapped), (6) Ac-Arg-NMe (dipeptide), (7) 6-(4-Morpholinyl)-2-pyridinecarbaldehyde, (8) Tris{2-[(dimethylamino)methyl]phenyl}silane, (9) Butyl 2-[(2,4-dichlorophenyl)amino]-4,4-dimethyl-6-oxo-1-cyclohexene-1-carbodithioate.
    Depicted elements are H (white), C (grey), N (blue), F (light green), O (red), Pb (dark grey), Cd (gold), Ba (dark green), Si (bronze), Cl (green) and S (yellow).}
\end{figure}

Figure~\ref{fig:molecules_app} shows example molecules from QM9, AA and OE and their chemical names.

\begin{figure*}[]
\includegraphics[width=18cm]{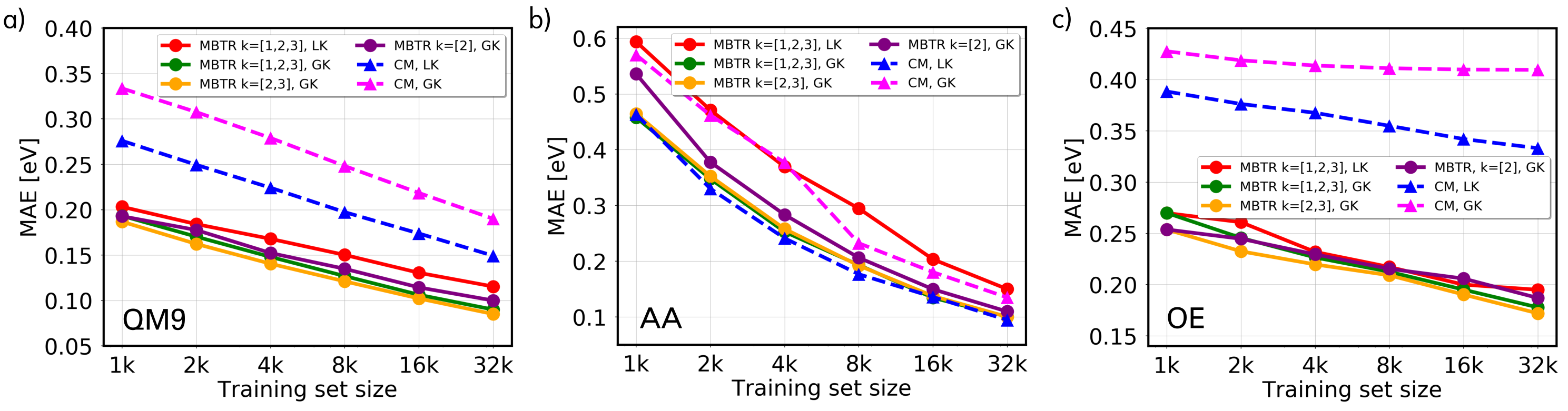}
\caption{\label{fig:comparison}Learning curves for different parameter configurations of the CM and the MBTR for the three datasets QM9, AA and OE. For the MBTR, the letter k refers to many-body terms, i.e. k=1 refers to one-body terms describing the atoms types that are present in a molecule, k=2 refers to two-body terms describing pairwise inverse distances between atoms and k=3 refers to three-body terms describing angle distributions for any triplets of atoms. Investigated combinations of many-body terms are one-, two- and three-body terms (k=[1,2,3]), two- and three-body terms (k=[2,3]) and only two-body terms. In addition, we consider two kernels, the Gaussian kernel (GK) and the Laplacian kernel (LK).}
 \end{figure*}

 \section{\label{sec:kernels}Choice of kernels and MBTR terms}
In Fig.~\ref{fig:comparison}, we show learning curves for different parameter configurations of the CM and the MBTR. MAEs correspond to out-of-sample predictions made in a single experiment. We generated learning curves using the MBTR with one-, two- and three-body terms (k=[1,2,3]), two-and three-body terms (k=[2,3]) and only two-body-terms (k=2). For QM9 and OE, we the MBTR with k=[2,3] performs slightly better than the MBTR with k=[1,2,3], while for AA, the performance of k=[2,3] and k=[1,2,3] is equal for larger training set sizes. Therefore, we chose to employ the MBTR with two- and three body terms (k=[2,3]) in this study. Moreover, Fig.~\ref{fig:comparison} reveals that, when the MBTR is used as molecular descriptor, the Gaussian kernel (GK) yields better results than the Laplacian kernel (LK). For the CM, on the other hand, the Laplacian kernel works better than the Gaussian kernel due to discontinuities of the sorted CM.

\section{Cross-validated grid search}\label{sec:cv}
For MBTR and KRR hyperparameter selection, we employed a 5-fold cross-validated grid search. In a 5-fold cross validation, a given \emph{original training set} (1k, 2k, 4k, 8k, 16k or 32k) is shuffled randomly and split into 5 equally sized groups. One group (20\% of the training set) is taken as a hold-out set for \emph{validation}. The remaining groups (80\% of the training set) are taken as \emph{training data}. Then, a grid search is performed: Each possible combination of KRR and MBTR hyperparameter values is trained on the \emph{training data} and evaluated on the \emph{validation data}.

The assignment of the 5 groups into \emph{training data} and \emph{validation data} is repeated 5 times, until each group was used as \emph{validation data} exactly once and used as \emph{training data} exactly 4 times. As we repeat the process 5 times, we get 5 MAEs for each possible set of hyperparameters. We consider the average value over these 5 MAEs for each set of hyperparameters and choose the set with lowest average MAE. With the chosen set of optimal hyperparameters, we train the KRR model on the entire \emph{original training set} (=\emph{training data}$+$\emph{validation data}). Finally, we evaluate the trained model on the test set of 10k out-of-sample molecules and report the MAE on the test set in the final learning curve for the \emph{original training set} size.

\nocite{*}

%

\end{document}